%% file: paper.tex
\documentclass[11pt,a4paper]{article}

\usepackage[a4paper,margin=2.6cm]{geometry}
\usepackage{amsmath,amssymb}
\usepackage{graphicx}
\usepackage{booktabs}
\usepackage{array}
\usepackage{longtable}    
\usepackage{xcolor}
\usepackage{caption}
\usepackage{microtype}
\usepackage[section]{placeins}   
\usepackage[numbers,sort&compress]{natbib}
\usepackage[colorlinks=true,linkcolor=blue!50!black,citecolor=blue!50!black,
            urlcolor=blue!50!black]{hyperref}

\graphicspath{{figures/}}
\newcommand{\Nsig}{N_{\mathrm{trials}}}
\newcommand{\Zloc}{Z_{\mathrm{local}}}
\newcommand{\Zglob}{Z_{\mathrm{global}}}
\newcommand{\TeV}{\ensuremath{\,\mathrm{TeV}}}
\newcommand{\GeV}{\ensuremath{\,\mathrm{GeV}}}
\newcommand{\fbinv}{\ensuremath{\,\mathrm{fb}^{-1}}}
\newcommand{\budgetvalues}{the values used throughout}

\title{\bfseries The Search Budget of the BSM Resonance Program}
\author{%
  Theresa Reisch\thanks{\texttt{theresa.reisch@unige.ch}} \qquad
  Tobias Golling\thanks{\texttt{tobias.golling@unige.ch}} \\[6pt]
  \normalsize D\'epartement de Physique Nucl\'eaire et Corpusculaire, Universit\'e de Gen\`eve \\
  \normalsize 24 Quai Ernest-Ansermet, 1211 Gen\`eve 4, Switzerland
}
\date{}

\begin{document}
\maketitle

\begin{abstract}
\noindent
Data-directed scans, anomaly searches and general searches probe many mass spectra at once, and for
them the look-elsewhere effect sets the discovery bar. We count the trials factor of the ATLAS BSM
resonance program in effective independent resolution elements, from public inputs alone, and then
work out what a fully combinatorial scan would cost in the same units. The published record, summed
over the 104 spectra it scans, amounts to $\Nsig = 7.9\times10^{3}$ looks, so a $5\sigma$ global
discovery today costs a local $\Zloc = 6.55$.
Scanning instead every mass built from up to four reconstructed objects, one event-level selection
at a time and priced on Run~2 and Run~3 together, amounts to $3.6\times10^{5}$ looks and
$\Zloc = 7.11$. A factor $46$ more looks therefore costs $0.56\sigma$. If the trials factor cannot
be enumerated because the estimator is itself
imperfect, as in machine-learning searches, two-stage unblinding is a safeguard against its spurious
signals: at the defect rate measured for a published bump-hunting network it is the more sensitive
procedure.
\end{abstract}

\section{Introduction}
\label{sec:intro}

A conventional search fixes model, final state and mass window before looking at data; the
look-elsewhere effect is then a small correction confined to that search. Data-directed
scans~\cite{ddp,bumpnet}, machine-learning anomaly searches~\cite{atlas_ml_anomaly} and general
searches over many event classes~\cite{atlas_general_search} instead sweep large parts of the mass
space at once, and for them the trials factor dominates the sensitivity. The total number of
independent looks in the BSM resonance program has nevertheless never been counted: it is quoted
per analysis at best, field-wide only as order-of-magnitude
estimates~\cite{diggingdeeper,statinterp}.

Throughout, one \emph{search} is one scanned mass spectrum, $n_s$ is its number of effective
independent resolution elements, and $\Nsig = \sum_s n_s$. In the fixed-resolution-element
approximation of the asymptotic Gross--Vitells relation~\cite{grossvitells}, the local significance
required for a $5\sigma$ global discovery~\cite{lyons2008} is
\begin{equation}
  \Zglob^2 \;=\; \Zloc^2 - 2\ln \Nsig
  \qquad\Longrightarrow\qquad
  \Zloc \;=\; \sqrt{25 + 2\ln \Nsig}\,.
  \label{eq:lee}
\end{equation}

Table~\ref{tab:granularity} collects the counts, a ladder from the narrowest defensible one to the
widest. Two of its rows do most of the work: the searches ATLAS has published cost
$\Zloc = 6.55$, and scanning every mass the detector can build costs $7.11$. What separates them is
how much of the mass space is swept, and Eq.~\eqref{eq:lee} compresses the whole ladder, a factor 84
in $\Nsig$, into $0.65\sigma$. Each bar carries a systematic band of two to three tenths of a sigma,
taken
input by input in Appendix~\ref{app:uncert}. We consider three bases: the spectra the model
space requires (Section~\ref{sec:budget}), the searches ATLAS has published
(Section~\ref{sec:published}), and a hypothetical combinatorial scan (Section~\ref{sec:scaled}).
Section~\ref{sec:counting} states the rules that turn the spectra into a number of looks.
Sections~\ref{sec:ab} and~\ref{sec:selection} then treat the case in which the trials factor cannot
be
enumerated at all, because the estimator taking the looks is itself imperfect. Every input is
public,
and every number, table and figure here is reproduced by the repository at
\url{https://github.com/rodem-hep/search-budget}.

\begin{table}[htbp]
\centering
\caption{The trials count on every basis, with what each is counted from. Rows are counted
differently and are never summed: the spectra column counts each basis's own unit (axes, event
selections, charged (search, axis) pairs, or per-category histograms). Bands for the remaining rows
are in Appendix~\ref{app:uncert}.}
\label{tab:granularity}
\small
\begin{tabular}{llrrr}
\toprule
basis & counted from & spectra & $\Nsig$ & $\Zloc$ \\
\midrule
model space, inclusive & published scan windows & 63 & $4.3\times10^{3}$ &
  $6.46\,^{+0.18}_{-0.16}$ \\
model space, event selections & the slices published searches use & 111 & $7.2\times10^{3}$ & 6.54 \\
published ATLAS program & 86 census entries, own ranges & 104 &
  $7.9\times10^{3}$ & 6.55 \\
combinatorial scan, Run 2+3 & every fittable ${\le}4$-object mass & 4\,438 & $2.0\times10^{5}$ &
  $7.03\,^{+0.23}_{-0.31}$ \\
\quad with event-level selections & one selection lens at a time & 8\,211 & $3.6\times10^{5}$ &
  $7.11\,^{+0.24}_{-0.32}$ \\
\bottomrule
\end{tabular}
\end{table}

\section{Counting the Looks}
\label{sec:counting}

Two resonances in one spectrum are distinguishable if separated by about one mass resolution, so for
a constant fractional resolution $\sigma_M/M = r$ over a window $[M_{\mathrm{lo}}, M_{\mathrm{hi}}]$
the number of effective independent looks is
\begin{equation}
  n_s \;=\; \frac{1}{r}\,\ln\!\frac{M_{\mathrm{hi}}}{M_{\mathrm{lo}}}\,,
  \label{eq:ns}
\end{equation}
summed over disjoint segments where a spectrum is scanned in pieces. Two rules fix what goes into
it.

\paragraph{The mass resolution is the only physics input.} Each value is propagated from the ATLAS
performance measurement of the object that limits the mass resolution: calorimeter
$e$/$\gamma$~\cite{atlas_egamma_calib}, muon sagitta~\cite{atlas_muon_perf}, light- and $b$-jet
energy
resolution~\cite{atlas_jer, atlas_btag_perf}, hadronic-tau visible products~\cite{atlas_tau_reco},
large-radius jet mass for boosted $V$, $h$, $t$, $H$~\cite{atlas_largeR_mass}, and missing
transverse
momentum~\cite{atlas_met_perf}. Since these are performance papers for the object, and none quotes a
resolution for a particular bump hunt, $r$ is known only to a factor of a few: scaling it by two in
each direction is worth $\pm0.11\sigma$ on the bar, and only $^{+0.02}_{-0.02}$ if the per-channel
errors are independent rather than common (Appendix~\ref{app:uncert}). Appendix~\ref{app:resolution}
documents each value, the only place a constant $r$ is a real approximation, and the
narrow-resonance
assumption behind counting resolution elements at all. The resolution stays a detector property only
as long as the background model is an empirical fit tuned channel by channel. A network returning
local
significances across a whole family of histograms~\cite{ddp,bumpnet}, or nonparametric background
estimation with Gaussian-process regression~\cite{gp_rw,gpr_frate,gpr_barr}, drives the marginal
cost
of one more spectrum towards zero, and would make a scan of the width priced in
Section~\ref{sec:scaled} practical. Both replace $r$ by an effective resolution, the network's or
the
kernel length scale, which smooths over a narrow bump when long and absorbs it when short, so a scan
built on either has to measure $\Nsig$ by toys on background-only data rather than read it off
Eq.~\eqref{eq:ns}.

\paragraph{The definition of one look is a convention, and the leading uncertainty.}
Eq.~\eqref{eq:ns} counts elements one resolution wide and Eq.~\eqref{eq:lee} treats each as an
independent test; both approximate the expected number of excursions above the claim level, and they
differ by a factor of order unity that no measurement fixes. Correlated neighbours argue for fewer
looks, a resonance of width $\sigma_M$ lifting more than one element. Rice's formula argues the
other
way~\cite{rice1944,grossvitells}: for a smooth unit-variance Gaussian process in $x = \ln M$ with
correlation
length $r$ it gives
$\langle N_Z \rangle = (2\pi)^{-1} r^{-1} \ln(M_{\mathrm{hi}}/M_{\mathrm{lo}})\, e^{-Z^{2}/2}$
up-crossings of level $Z$, the element count times $Z/\sqrt{2\pi} \approx 2.6$ once compared against
the Gaussian tail Eq.~\eqref{eq:lee} multiplies. Running the effective trials factor from
$0.5\,\Nsig$ to $\Nsig Z/\sqrt{2\pi}$ moves the bar by $^{+0.14}_{-0.11}\sigma$ on the model space
and $^{+0.15}_{-0.10}$ on the scan: the largest single
term of the budget, and common to every basis, so it cancels in every difference quoted here. The
element count is the one we quote, because a reader can rebuild it from a published window and a
resolution.

\paragraph{A hypothetical spectrum has to hold enough events to fit.} A histogram counts as fittable
if it holds at least $100$ events and at least $25$ resolution elements with one event or more: the
first is the least a parametric background fit of two or three parameters
needs~\cite{atlas_dijet_run2},
the second is set by the trials count itself, a window of fewer than 25 elements not being a scan in
any useful sense. Both are round numbers, and Appendix~\ref{app:yield} prices them. Wherever this
paper
counts a spectrum nobody has published, the yields come from a declared power-law background,
$n(m) \propto W\,m^{1-P}$ with $P = 7$, anchored at $10^{6}$ events per element at $1\TeV$ in the
light-jet pair spectrum and scaled by one factor per object type, each fixed on the published
symmetric channel of that type. Since $n(m)$ falls, the requirement truncates every window where an
element drops to one event and discards the spectrum if fewer than 25 elements remain. Published
windows are exempt throughout, because a published search demonstrates its own feasibility.

\paragraph{The dataset.} All yields are evaluated on Run~2 and Run~3 combined: $140\fbinv$ at
$13\TeV$~\cite{atlas_lumi_run2} and roughly twice that again expected at $13.6\TeV$, three times the
Run-2
dataset the anchor is set on, ignoring the rise in high-mass cross sections between the two
energies.
That under-fills the high-mass tail and admits fewer spectra; of the modelling choices made here it
is the only one that biases the scan towards being cheap. Luminosity buys breadth slowly: the
one-event mass grows as $\mathcal{L}^{1/(P-1)}$, so tripling the dataset extends each window by a
fifth in mass.

\paragraph{Reinterpretation carries no trials cost.} Models are grouped by the spectrum the bump
appears in, because all models peaking in one spectrum are tested by one search: adding a model to
an already-scanned
spectrum costs zero trials, opening a new spectrum costs its $n_s$.

\section{The Model-Space Budget}
\label{sec:budget}

\begin{figure}[t]
\centering
\includegraphics[width=0.8\textwidth]{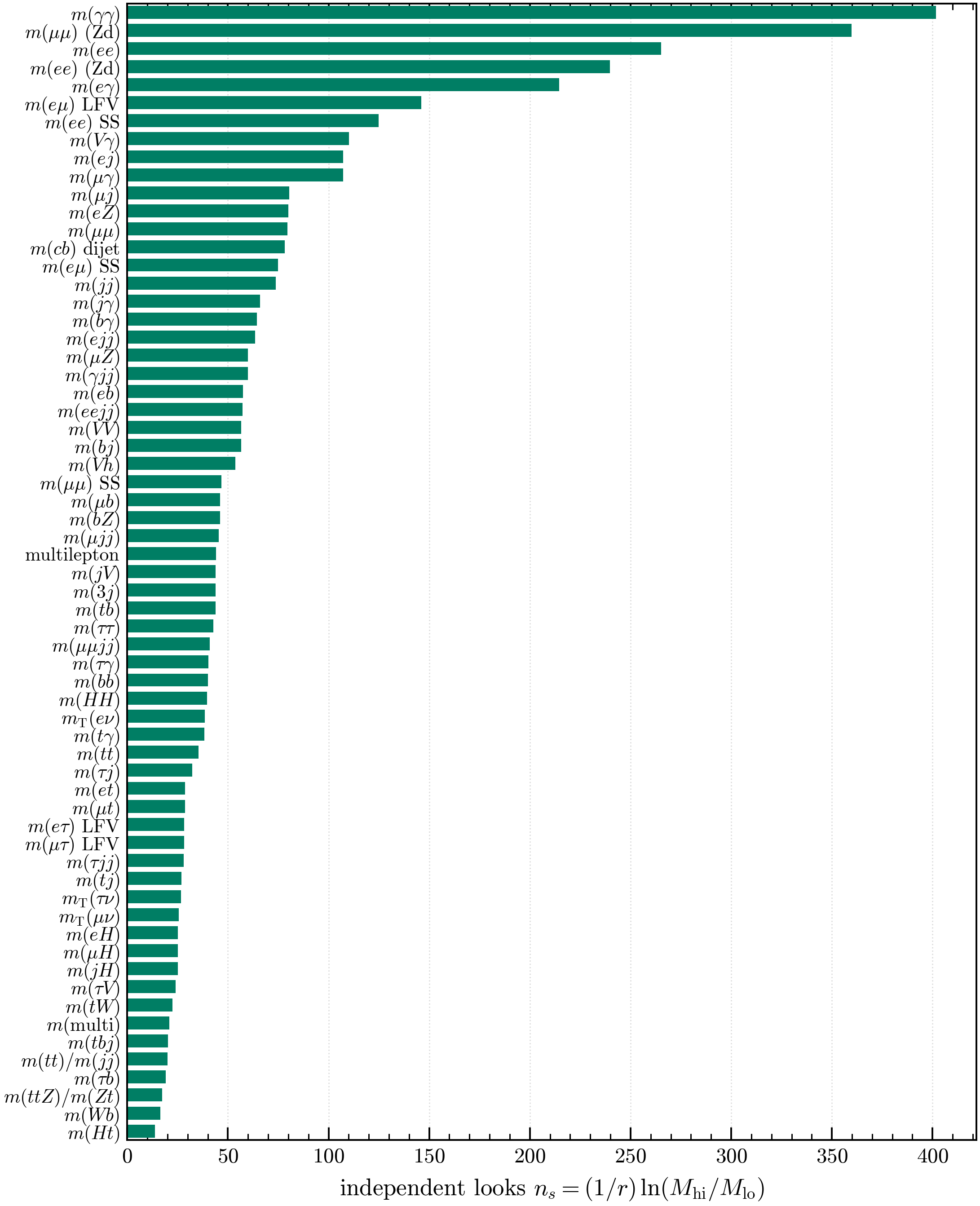}
\caption{Independent looks per bump spectrum over its published scan window.}
\label{fig:budget}
\end{figure}

The first basis is the model space: every spectrum in which some public BSM model predicts a peak,
whether or not a search for it exists. The public model classes collapse onto 63 spectra
(Figure~\ref{fig:budget}), with $m(VV)$, $m(jj)$ and $m(t\bar t)$ each motivated by nine to
thirteen classes, which is why there are far fewer searches than models. Table~\ref{tab:modelmap} in
Appendix~\ref{app:modelmap} is that map axis by axis, assembled from the FeynRules model
database~\cite{feynrules,feynrules_db}, the published search record and a sweep of the
resonance-model literature beyond the database, and is the input this whole
count is built from; the fifteen classes the database does not implement are cited individually
in the table, and a class with no reference behind it comes from the database, covered by that
single reference~\cite{feynrules_db}.
Individual spectra range from 14 looks for $m(Ht)$ to 402 for $m(\gamma\gamma)$, and the total is
$\Nsig = 4\,319$ over published scan windows, so $\Zloc = 6.46\,^{+0.18}_{-0.16}$
(Appendix~\ref{app:uncert}). Because $n_s \propto 1/r$ the sharpest axes carry the most, the five
largest contributors making $34\,\%$ of the total; one of them, $m(e\gamma)$, has been scanned only
at 7 and 8\TeV. No single spectrum dominates, and removing the largest, $m(\gamma\gamma)$, leaves
$6.45$.

One spectrum per observable is a lower bound, since searches also slice by $b$-tag category, boosted
against resolved, sub-decay mode, lepton multiplicity and pairing ambiguity: a publication-anchored
multiplicity gives 111 channels (Table~\ref{tab:modelmap}) and moves the bar to 6.54. Because
$\Nsig$
enters Eq.~\eqref{eq:lee} only through its logarithm, the convention is almost immaterial: inclusive
spectra, event selections and a kinematic envelope up to $\sqrt{s}$ agree to within a tenth of a
sigma.
Covering the entire known BSM resonance space costs about one and a half $\sigma$ over a
single counting experiment: $6.46$ against the $5.0$ a single look pays.

\section{The Published ATLAS Program}
\label{sec:published}

The second basis is the published record, which fixes the trials cost the program already pays and
can be tested against the excesses actually reported. A census assembled by hand groups 290 ATLAS
resonance papers\footnote{The 290 are not cited individually: a bibliography of that size would
outweigh the argument it supports. The complete record is committed in the
repository instead, in the form described in Appendix~\ref{app:census}.}
published between 2010 and 2026 into 86 catalogued searches over 64 bump observables
(Appendix~\ref{app:census}). It counts analyses rather than model-motivated axes, so it is never
added to the 63, but it can be priced in the same currency. Every entry carries the axes it scans
and the ranges it scanned, and the
resolution-element rule applies to it unchanged. Since 20 of the 86 scan more than one axis, the
entries make 114 (search, axis) pairs, of which 104 carry a mass range that can be charged
(Appendix~\ref{app:census}); charging each of those over its own range, the axis's union window
standing in where an entry scans a counted axis without quoting one, gives $\Nsig = 7\,875$
($\Zloc = 6.55$), while counting each axis once over the union of every published range on it gives
$3\,837$ ($6.44$).
The second sits within $11\,\%$ of the $4\,319$ the model space implies: the union basis misses the
nineteen model-space axes no published search has scanned and adds ten outside the 63, two gaps that
largely cancel. What the comparison supports is agreement at the ten-percent level, and that much
is not circular: the model space takes its windows from the published searches but its axes from the
models, so the agreement says the program scans almost exactly the axes the models motivate. Read
by date, 6 of the 86 have a published Run-3 result and 19 have no paper since 2019. Fourteen of
those 19 scan an axis already counted in $\Nsig$, so on the union basis revisiting them is free (the
per-search basis charges every re-scan); the other five, among
them the three-photon and triboson spectra, would extend the count like any new spectrum.
Counted instead in the units of the combinatorial scan of Section~\ref{sec:scaled}, where the
pair-produced axis resolves into its $tt$ and $jj$ legs per category, the models point at 21
spectra no published ATLAS search has scanned, 19 of them within the scan's reach
($m(\mathrm{multi})$ has no fixed object composition, and the $tt$ leg in the all-top category
fails the statistics requirement); the repository lists them leg by leg.

The $7\,875$ is a direct count where the literature has only estimates, and the two agree in order
of
magnitude: quasi-independent statistical tests across all ATLAS searches, signal regions of counting
experiments included, come to $\sim5\times10^{4}$~\cite{diggingdeeper,statinterp}, a broader tally
than the mass scans counted here.

A trials count of this size predicts the excess population the program should have produced. With
$p(\ge3\sigma) = 1.35\times10^{-3}$ and $p(\ge5\sigma) = 2.87\times10^{-7}$, $\Nsig p$ over the
census expects $11$ local $3\sigma$ fluctuations and $0.002$ spurious $5\sigma$. Both are what the
record holds: the LHC BSM Working Group lists about twenty open ATLAS
excesses above $2.4\sigma$ local, the largest at $3.5\sigma$ local and none above $2.8\sigma$
global~\cite{lhcbsmwg}; the abstracts of the 290 census papers report six local excesses at or above
$3\sigma$, against the $6$ a background-only sweep of the 63 axes expects and the $11$ it expects
over the census; and every ATLAS
$\ge5\sigma$ observation to date corresponds to a known Standard Model
process~\cite{atlas_public_searches}. Expected and observed
agree without adjustment, though the observed side is loose: abstracts under-report small excesses
and an open-excess list holds only what has not faded. It follows that over the published record a
local $5\sigma$ in a fully agnostic scan is worth $2.7\sigma$ globally, and over the lensed
combinatorial scan of Section~\ref{sec:scaled} nothing at all, which is the basis of the
proposal~\cite{statinterp} to raise agnostic-scan thresholds towards $7\sigma$; that scan reaches
essentially the same number from the trials side.

\section{The Cost of a Fully Combinatorial Scan}
\label{sec:scaled}

The third basis is hypothetical and the widest: what it would cost to build every invariant mass
from
every combination of reconstructed objects. Three terms are kept distinct. An \emph{axis} is a mass
observable, for instance $m(ee)$; a \emph{category} is an exclusive event class fixed by its object
multiplicities; a \emph{spectrum} is one axis in one category, the histogram that would be fitted,
so
one axis in twenty categories costs twenty times its looks.

\paragraph{The scan.} Reconstruction today provides ten object types that can enter a mass:
electrons, muons, hadronic taus, photons, light jets, $b$-jets, and the boosted large-$R$ candidates
for $W/Z$, $H$ and top, plus a leptonic $Z$. Categories are exclusive multiplicity bins of at most
four objects, as in the ATLAS general search~\cite{atlas_general_search}. No trigger is required,
which overfills the all-jet channels at low mass and errs towards more looks; missing energy splits
every category but never enters a mass, leaving the three transverse-mass axes ($W'$-type signals)
out of reach; and same-flavour dilepton categories
split by charge. Every size-2 to size-4 subset of the indexed
objects is its own spectrum, so $j_0j_1$, $j_0j_2$ and $j_1j_2$ count as three. Subsets sharing an
object are correlated tests on the same events, as are the lensed views below, so charging each as
an independent look over-counts; that only raises the bar, and raises the scan's more than any
published basis's, so the differences quoted between them are upper bounds. Each spectrum takes
$r = \tfrac12\,(\overline{\sigma^2})^{1/2}$ over its object group, with each object's $\sigma$
inverted from the published resolution of its own symmetric channel via $\sigma = 2r$, so no new
resolution input enters. The prefactor is calibrated to the symmetric channels rather than
propagated: it makes a four-body mass as sharp as a two-body one, and the worst-leg alternative
costs $-0.10\sigma$ (Appendix~\ref{app:uncert}).

That alphabet allows 21\,644 combinations in 2\,412 categories. The statistics requirement of
Section~\ref{sec:counting} removes four of every five, mostly at high multiplicity, leaving
4\,438 fittable spectra, $\Nsig = 2.0\times10^{5}$ and $\Zloc = 7.03\,^{+0.23}_{-0.31}$.
Two-body groups are $73\,\%$ of what survives, and the costliest compositions, $\gamma\gamma$,
$e\gamma$, $ee$ and $ej$, are the ones built from the sharpest objects. The band is wider than on
the model space because the yield model enters too, and it dominates: scaling its anchor by $10^{2}$
either way moves $\Zloc$ over $6.88$ to $7.14$, against $0.01$ for the background exponent over $6$
to
$8$ and $^{+0.04}_{-0.12}$ for the fittability thresholds (Appendix~\ref{app:uncert}).

\paragraph{What the scan adds.} A spectrum here is one axis in one category. We ask two
things of the $4\,438$: how many a public model predicts, and how many of those have never been
scanned. Appendix~\ref{app:newspectra} works out both. $164$ of the $4\,438$, or $3.7\,\%$, sit in a
final state some model produces, with the resonance on the sub-system the mass is built from. $51$
of those sit on an
axis no published ATLAS search scans, and they carry $2\,329$ looks over 18 axes. That set is what a
combinatorial scan adds to the published program. The rest, $96\,\%$ of the spectra and
$1.9\times10^{5}$ of the looks, is territory no published model points at, and it is what takes the
bar from $6.46$ to $7.03$.

\paragraph{Selection lenses on the same axis.} A wide search would also re-scan each axis under
event-level requirements: a high $H_T$ or $M_\mathrm{eff}$ threshold, displaced activity, a forward
jet pair for vector-boson fusion, an ISR jet for trigger acceptance at low mass. Missing energy,
multiplicity and $b$/$\tau$ enrichment are already inside the category count and would be double
counted. Appendix~\ref{app:yield} prices the four conservatively, one at a time and never a product
of
two, with axis, resolution and window unchanged and each view paying the statistics of its own
efficiency. Statistics rules out three of every five possible views, and the survivors take the scan
to 8\,211 histograms and $\Zloc = 7.11$, so a lens costs $0.08\sigma$ on the bar plus the events it
discards. That $0.08$ is a difference between two rows counted alike and better determined than
either: moving every declared efficiency by an order of magnitude leaves the lensed bar in
$7.08$ to $7.15$.

With the lenses added, something qualitative changes as well. Eq.~\eqref{eq:lee} loses meaning once
$25 - 2\ln\Nsig < 0$, at $\Nsig = e^{12.5} \approx 2.7\times10^{5}$. The unlensed scan sits below
that
line, a local $5\sigma$ still carrying $0.8\sigma$ globally; the lensed scan crosses it, a local
$5\sigma$ is worth nothing globally, and $5\sigma$ global demands $7.1\sigma$ local, the threshold
recommendation of Ref.~\cite{statinterp} reached from the trials side.

\paragraph{The two-body layer is a finite list.} At $K=2$ the gap can be named rather than
estimated.
Kim, Kong, Nachman and Whiteson~\cite{twobody} mark which pairs of decay products a published search
covers; Table~\ref{tab:twobody} replaces their marks by prices. Sixteen pairs carry no published
ATLAS scan, and twelve of them the model catalogue already holds as an axis -- among them
lepton-plus-top for the light flavours, $\tau V$, jet-plus-boson and jet-plus-Higgs -- so their
looks already sit inside $\Nsig$. Four have no axis at all, and closing them costs 151 looks,
$\Nsig$ going from $4.32$ to $4.47\times10^{3}$ and $\Zloc$ from $6.46$ to $6.47$: one hundredth
of a sigma, for a finite list of four spectra rather than a hypothetical program.

An upright cell in that table is what the published program already spends on the pair, summed over
every canonical axis a published search scans, each over its own published window at its own
resolution; the three $ee$ axes sum to 630, the most heavily scanned pair. An italic cell in
parentheses is a pair no published ATLAS search scans, priced at what one axis there costs. For
twelve of them the catalogue holds that axis -- the empty pub.\ column of
Table~\ref{tab:modelmap} -- so their looks are counted at their catalogue windows and already sit
inside $\Nsig = 4{,}319$.

The other four have no catalogue axis: with no window or resolution to
take, they are priced alike, on the $100\GeV$--$5\TeV$ window and the pair resolution
$r = \tfrac12\,(\overline{\sigma^2})^{1/2}$ of the combinatorial scan, truncated like any
hypothetical spectrum, running from 30 looks ($\tau H$) to 52 ($\gamma H$); closing those four is
the $+151$ above. Three of them, $\tau t$, $bH$ and $\gamma H$, are pairs ATLAS has in fact
published a scan for -- the leptoquark $t\tau$ leg, the single vector-like-quark $Hb$ mode and the
$H\gamma$ resonance -- without a public model class in the catalogue pointing at them, so each is
one of the ten published observables outside the 63 rather than a budget axis. The fourth, $\tau H$,
is motivated but fails the statistics requirement on its own kinematic floor, holding 24.6
resolution elements against the 25 the yield model asks for. Two scanned pairs
also hold an unscanned axis each, the charm-tagged dijet in $bj$ and the coloron-pair leg in $tt$,
counted in the budget but not in the upright number. The two kinds of cell are not comparable one to
one, since an upright cell may hold several axes and published windows are mostly narrower, so a
small upright number says the published window is narrow and nothing about how cheap the pair is.

\begin{table}[htbp]
\centering
\caption{The two-body grid of Ref.~\cite{twobody} priced in trials. Upright: looks a published ATLAS
search already spends on that pair. Italic, in parentheses: no published search scans the pair, and
the looks one axis there costs.}
\label{tab:twobody}
\small
\input{two_body_matrix}
\end{table}

\section{A Safeguard Against Spurious Signals}
\label{sec:ab}

Every count so far was enumerable because published windows and resolutions can be looked up. A
network-driven scan offers neither, and it adds a problem the Gaussian arithmetic does not cover:
the significance estimator itself is imperfect. For one estimator of this kind the failure rate has
been published. BumpNet~\cite{bumpnet}, a network predicting per-bin significances across families
of mass
histograms, flags above $5\sigma$ on background-only inputs in $0.048\,\%$ of histograms built from
smooth analytical functions and $0.129\,\%$ of those built from simulated Standard Model samples;
applied to its $39\,746$ application histograms with no signal injected it returns $53 \pm 7$
spurious
$5\sigma$ candidates, all four numbers as published there. These are not fluctuations: at $15$ to
$60$ elements per histogram the set holds
on the order of $10^{6}$ looks, for which the Gaussian expectation is below one. The measured rate,
$10^{-5}$
to $10^{-4}$ per look, sits one to two orders of magnitude above the statistical one and is a lower
bound on the fraction $\epsilon$ of looks whose significance is simply wrong, since only
mis-estimates reaching $5\sigma$ are counted; at the $3\sigma$ level where candidates are selected,
$\epsilon$ may well reach $10^{-3}$.

\begin{figure}[t]
\centering
\includegraphics[width=0.92\textwidth]{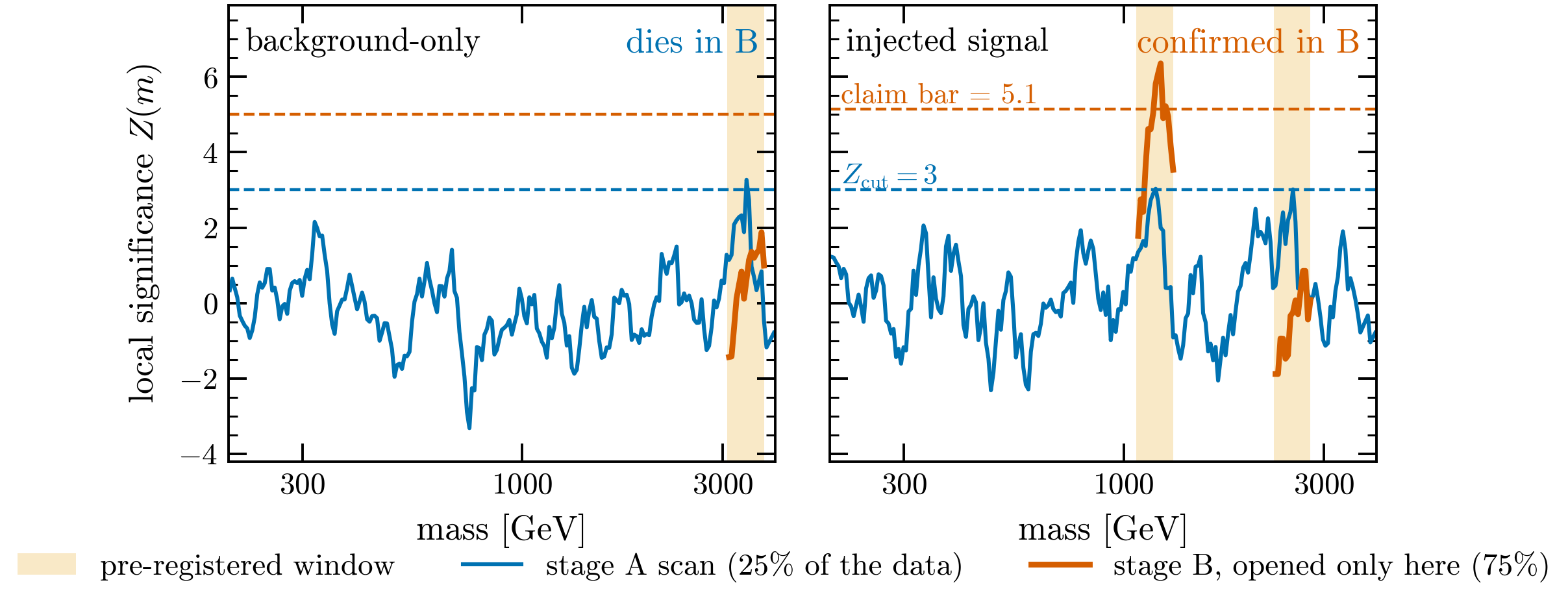}
\caption{The two-stage procedure on a toy falling spectrum ($5\,\%$ resolution, $f = 0.25$). Left: a
$3.3\sigma$ fluctuation is pre-registered and stage B shows $0.6\sigma$. Right: an injected
$Z_{\mathrm{full}} = 7$ signal confirms at $6.4\sigma$, above the claim bar.}
\label{fig:abguard}
\end{figure}

For a single-stage scan this poisons the correction itself: Eq.~\eqref{eq:lee} calibrates the tail
of
$\Nsig$ Gaussian looks and the measured tail is about $10^{2}$ times heavier, so the effective
trials
factor becomes a property of the network that nobody can enumerate afterwards. Splitting the dataset
converts that inflation into a countable list. Explore a fraction $f$ of the data, pre-register
every
window whose stage-1 significance exceeds $Z_{\mathrm{cut}}$, and unblind the remaining $1-f$ only
there, as the BumpNet authors propose for LHC data~\cite{bumpnet}. The confirmation stage then
carries
the trials factor of the frozen list, whatever produced the entries: a claim bar
$\sqrt{25 + 2\ln(wk)}$ for $k$ pre-registered windows, each allowed $w$ resolution elements of mass
freedom, $w = 3$ throughout (about $\pm2\sigma_M$). And the list cleans itself: a spurious entry,
fluctuation or artefact
redrawn at each evaluation, does not repeat in the independent confirmation data
(Figure~\ref{fig:abguard}). On the combinatorial scan of Section~\ref{sec:scaled},
$\Nsig = 3.6\times10^{5}$, a cut at $Z_{\mathrm{cut}} = 3$ pre-registers $490$ windows per
experiment,
and in $2\times10^{4}$ background-only toys the best confirmation never comes close to the claim bar
($5.30$ against $6.29$), so not one false claim.

The safeguard is not free: at the optimum ($f \approx 0.2$--$0.3$,
$Z_{\mathrm{cut}} \approx 2$--$3$)
the split reaches $50\,\%$ discovery power at $Z_{\mathrm{full}} \approx 7.6$ where an exactly
corrected
single-stage scan of the same spectra needs $7.11$. The price is $+0.47\sigma$ at the $w = 3$ mass
freedom used here and $+0.32$ with pinned windows, and it barely depends on the
trials count it is paid on: a factor two in $\Nsig$ changes it by $0.01\sigma$. That insensitivity
also fixes where the two procedures trade places. The split is the more sensitive one whenever the
trials factor a single-stage analysis would honestly have to defend exceeds about $14$ times the
counted one: implausible for a classical fixed-window scan, but an order of magnitude below the
$10^{2}$ measured for a BumpNet-class estimator, where the split comes out ahead on sensitivity
alone, before auditability even enters the argument. That comparison charges the single stage the
full
measured inflation; upstream defect suppression, like the correlation veto below, narrows the
margin,
but would have to remove a factor of six before the verdict changes. The split itself is hurt
only logarithmically: if that inflation persists at $Z_{\mathrm{cut}}$, the expected list grows from
$4.9\times10^{2}$ to $4.9\times10^{4}$
entries and the claim bar from $6.29$ to $6.99$, and even that assumes the length is modelled rather
than read off the
frozen list.

Requiring both stages to succeed gives the joint discovery probability
\begin{equation}
  P_{\mathrm{disc}} = \Phi\!\left(\sqrt{f}\,Z_{\mathrm{full}} - Z_{\mathrm{cut}}\right)
  \Phi\!\left(\sqrt{1-f}\,Z_{\mathrm{full}} - Z_B^{\mathrm{req}}\right),
  \qquad
  Z_B^{\mathrm{req}} = \sqrt{25 + 2\ln (w\,k_{\mathrm{eff}})},
  \label{eq:abpower}
\end{equation}
with $k_{\mathrm{eff}} = \Nsig\,p_1(Z_{\mathrm{cut}}) + 1$ the expected number of pre-registered
windows and $w$ the mass freedom per window as above. The reach is the $Z_{\mathrm{full}}$ at which
$P_{\mathrm{disc}} = 50\,\%$, and toy Monte Carlo
reproduces it to $0.03\sigma$. No choice of $f$ closes the penalty at any trials factor: the
look-elsewhere effect factorises, $\Nsig = (\Nsig/k)\cdot k$, so moving the looks into two stages
destroys none of them, while the split luminosity costs $\sqrt{2}$ on $Z$ and both stages must now
succeed. The cost is markedly asymmetric between the two stages, small exploration and large
confirmation, so the naive $50/50$ split is a poor choice (Table~\ref{tab:abcost}).

\begin{table}[htbp]
\centering
\caption{Signal strength required for a $50\,\%$ probability of discovery, over the
$\Nsig = 3.6\times10^{5}$ independent looks of the combinatorial scan, at $w = 3$ elements of mass
freedom per pre-registered window.}
\label{tab:abcost}
\begin{tabular}{lrr}
\toprule
procedure & reach $Z_{\mathrm{full}}$ & cost \\
\midrule
single stage on the full dataset, exactly corrected & 7.11 & -- \\
optimised split ($f \approx 0.2$--$0.3$, $Z_{\mathrm{cut}} \approx 2$--$3$) & $\sim7.6$ & $+0.5\sigma$ \\
naive $50/50$ split, $Z_{\mathrm{cut}} = 3$ & $\sim8.9$ & $+1.8\sigma$ \\
\bottomrule
\end{tabular}
\end{table}

One class of defect survives the split: a coherent one. Splitting suppresses what fails to repeat,
and
a mismodelled background or a detector artefact that scales with luminosity as a signal would passes
confirmation as readily as a real signal. The BumpNet application shows the class is real: after the
veto
on cross-histogram correlations reported in Ref.~\cite{bumpnet}, $27$ of the $53$ candidates
survive, a coherent fraction of
$51 \pm 7\,\%$, clustered at the start of their histograms, shape-driven rather than statistical.
That
fraction is worth measuring before any unblinding: evaluate the estimator on two independent
background-only samples under identical histogram definitions and count how often a flag lands in
the
same window twice.

\section{Selecting Candidates with an Imperfect Estimator}
\label{sec:selection}

Section~\ref{sec:ab} forwarded every window above an absolute $Z_{\mathrm{cut}}$; that choice
matters more than any counting convention, and the measured defect rate decides it. Three rules
are in use: take the largest bin (\emph{argmax}), take every bin above an absolute threshold $t$,
or apply a Benjamini--Hochberg false-discovery-rate rule~\cite{benjaminihochberg} at nominal $q$.
A fair comparison holds the false-alarm budget fixed; we set it to $1.35\times10^{-3}$ expected
false confirmations, the rate the argmax produces by construction.

Under a perfectly calibrated estimator the threshold confirms more signal than the argmax at every
signal strength, with BH between them, and the reason is structural: on a common ROC the threshold
traces a
curve as $t$ varies while the argmax, always returning exactly one bin, is a single point strictly
inside it. The margin is $+4$ percentage points at $\mu = 5\sigma$. The realistic regime is the
contaminated one. The two \textsc{glitch} rows of Table~\ref{tab:selnoise} carry a defect redrawn at
each evaluation, at rates running from the one measured for BumpNet at its $5\sigma$ flag level to
the
$10^{-3}$ it may reach at the $3\sigma$ level where candidates are selected (Section~\ref{sec:ab});
the \textsc{bias} row carries one that repeats. The rules are not equally exposed: argmax and BH
are rank-based, so an artefact anywhere changes the decision about every other bin, while a
fixed threshold is absolute and an artefact at $1\TeV$ says nothing about the bin at $3\TeV$.
The comparison therefore sharpens as the estimator degrades, the threshold's margin over the argmax
widening from $+4$ to $+8$ or $+13.6$ points, and a rank-1 rule degrades abruptly: the
probability that the argmax \emph{is} the signal, before any confirmation, falls from $80\,\%$ to
$45\,\%$, since any one of a few
dozen artefact bins outranking it removes the signal entirely, whereas a threshold pays only through
the extra candidates it forwards.

\begin{table}[htbp]
\centering
\caption{Probability that a $5\sigma$ signal is confirmed, each rule tuned to the same
$1.35\times10^{-3}$ false-confirmation budget. BH carries a Monte Carlo error of $\pm0.3$ points,
the other two columns none.}
\label{tab:selnoise}
\begin{tabular}{lrrrr}
\toprule
estimator & argmax & threshold & BH & thr $-$ argmax \\
\midrule
perfect & 78.5\,\% & 82.5\,\% & 80.9\,\% & $+4.0$ \\
\textsc{glitch} $\epsilon = 10^{-4}$ & 73.2\,\% & 81.2\,\% & 80.1\,\% & $+8.0$ \\
\textsc{glitch} $\epsilon = 10^{-3}$ & 43.8\,\% & 57.3\,\% & 57.2\,\% & $+13.6$ \\
\textsc{bias}, any $\epsilon$ & \emph{untunable} & $\sim0\,\%$ & \emph{unreachable} & -- \\
\bottomrule
\end{tabular}
\end{table}

\begin{figure}[htbp]
\centering
\includegraphics[width=0.56\textwidth]{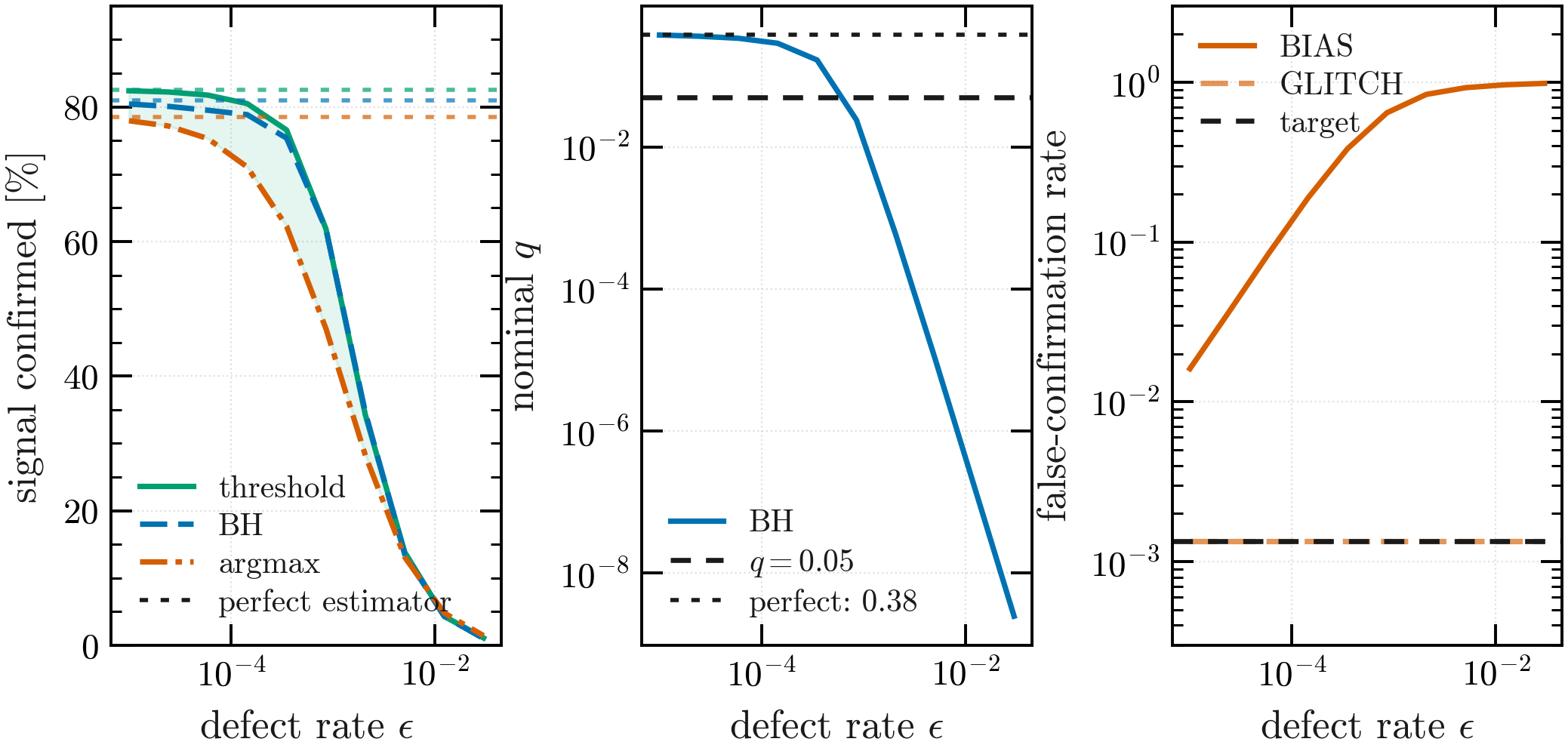}
\caption{Left: confirmation probability for a $5\sigma$ signal as the estimator degrades, all rules
at the same false-alarm budget (dotted: perfect estimator). Middle: the nominal FDR that buys that
budget. Right: the false-alarm rate the argmax achieves.}
\label{fig:selnoise}
\end{figure}

Contamination also strips BH of its adaptivity: to hold the budget it must run at ever smaller $q$,
and
BH at small $q$ \emph{is} a threshold at $\Phi^{-1}(1-q/n)$, converging onto the fixed rule while
its $q$ loses any useful interpretation. The nominal $q$ that buys the budget runs from $0.38$
(perfect) through
$0.32$ and $1.4\times10^{-2}$ across the \textsc{glitch} rows, and against a repeating defect cannot
be
bought at any $q$ (Figure~\ref{fig:selnoise}). None of this contradicts the FDR guarantee: BH
controls
the false discovery rate under the null it is handed, and a contaminated scan does not hand it that
null. The \textsc{bias} row is the coherent class of Section~\ref{sec:ab} and no selection rule
survives it; it must be removed upstream, not tuned away here. We therefore recommend selecting
stage-1 candidates on absolute significance at a fixed expected-fake budget, and quoting the
\emph{achieved} false-alarm rate rather than a nominal FDR parameter.

\section{Conclusions}

Organised by where the bump appears, the BSM resonance space is 63 spectra and
$\Nsig \approx 4.3\times10^{3}$ independent looks: a $5\sigma$ global discovery costs
$\Zloc = 6.46\,^{+0.18}_{-0.16}$, and $6.55$ for the published ATLAS program counted search by
search. Switching
convention, to event selections or a kinematic envelope, moves it by less than $0.1\sigma$ again,
and
the answer is consistent, without adjustment, with the observed ATLAS excess population.

On Run~2 and Run~3 together, a scan of every mass built from up to four
reconstructed objects costs $\Zloc = 7.03\,^{+0.23}_{-0.31}$, six tenths of a sigma above the known
model space, a difference known to $^{+0.12}_{-0.23}$, and event-level lenses add $0.08$ more. The
statistics requirement removes four of every five combinations before the look-elsewhere effect
weighs on any of them: a wide scan is limited by how many events a histogram holds, not by the
trials factor.

Width turns out to be the cheap part. The expense is scanning with an estimator whose mistakes
cannot be counted, and there two-stage unblinding converts the unmeasured trials inflation into a
countable list and kills every spurious signal that fails to repeat, at a price of half a sigma the
measured inflation
repays; stage-1 candidates should be selected by an absolute threshold rather than by rank; and the
coherent defect the split cannot kill is measurable before unblinding. The count is per experiment:
CMS scanning the same axes doubles the trials, moving the bar by $0.1\sigma$.

Appendix~\ref{app:uncert} puts numbers on the caveats: what one look \emph{is} costs
$^{+0.14}_{-0.11}$ to
$^{+0.15}_{-0.10}\sigma$ across the bases,
more than any physics input, and cancels in every difference; a factor two
on the resolutions $\pm0.11$; the
flat stand-in for the muon axes' rising $r(M)$ under $0.01$; a $40\,\%$ move of every window edge
$0.04$; and a model class missing from a counted spectrum nothing at all.

\section*{Acknowledgments}
This work is supported by the European Union's Horizon Europe research and innovation program under
the Marie Sk\l{}odowska-Curie grant agreement No 101168829 called ``Challenging AI with Challenges
from Physics: How to solve fundamental problems in Physics by AI and vice versa (AIPHY)''.

\bibliographystyle{unsrt}
\bibliography{refs,model_refs}

\clearpage
\appendix
\input{resolution_appendix}

\section{The Uncertainty Budget}
\label{app:uncert}

Every number here follows from a declared input: a resolution, a scan window, a yield anchor, a
fittability threshold, or the convention that turns a resolution element into an independent look.
Table~\ref{tab:uncert} moves each over the range it is honestly known to, re-running the count
rather
than scaling $\Nsig$ through Eq.~\eqref{eq:lee}: that matters for hypothetical spectra, where a
coarser resolution costs looks twice, through Eq.~\eqref{eq:ns} and through the 25-element
requirement a widened window now fails, which is why the scan's resolution band is the wider and
more asymmetric one.

Two features of the table matter more than any individual entry. The largest term comes from the
definition of one look, priced in Section~\ref{sec:counting}, and not from any physics input. And
every input except the yield model enters both bases alike, so the last column, the shift in the
difference, is small where
the first two are not: the resolution scale is worth $0.11\sigma$ on the model space and up to
$0.19$
on the scan but $^{+0.03}_{-0.08}$ on the gap, the look convention cancels to $0.01$, and only the
yield model, which the model space never uses, survives. Every statement of the form
``$X$ costs $\delta\sigma$ more than $Y$'' is therefore better determined than either bar.

Four choices are reported as alternatives rather than folded in, since they change what is counted,
not how well it is known: the granularity ($6.46$ against $6.54$), the dataset ($7.00$ for Run~2
alone
against $7.03$), the lens layer ($7.03$ against $7.11$) and the shape of the hypothetical scan
itself.
The per-channel resolution row is an alternative in the same sense, a second way of drawing the same
input rather than a source of its own, so a dagger marks it and the total leaves it out.
The event-selections row of Table~\ref{tab:granularity} carries the band of the row above it, the
two
differing by a multiplicity rather than an input; the published-program row is priced by the same
rule off the same resolutions, so it carries that band too, $6.45$ to $6.66$ under the resolution
scale alone; and the lensed scan carries $^{+0.24}_{-0.32}$.

\begin{table}[htbp]
\centering
\caption{Every declared input, the range it is varied over, and the shift in $\Zloc$ on the two
counted bases and on the difference between them. A dagger marks a second way of drawing an input
already listed, quoted against its own median and left out of the total.}
\label{tab:uncert}
\footnotesize
\renewcommand{\arraystretch}{1.35}   
\input{uncertainty_table}
\end{table}

\section{The Yield Model}
\label{app:yield}

The statistics requirement of Section~\ref{sec:counting} applies to the combinatorial scan of
Section~\ref{sec:scaled} and to the four two-body pairs with no catalogue axis priced there. Every
background here is a
steeply falling mass spectrum, so the events in the resolution element at mass $m$ are taken as
\begin{equation}
  n(m) \;=\; N_{\mathrm{ref}}\; W \left(\frac{m}{M_{\mathrm{ref}}}\right)^{1-P}
             \frac{r}{r_{\mathrm{ref}}}\,,
  \qquad
  N_{\mathrm{ref}} = 10^{6},\;\; M_{\mathrm{ref}} = 1\TeV,\;\; r_{\mathrm{ref}} = 0.05,\;\; P = 7,
  \label{eq:yield}
\end{equation}
anchored on the light-jet pair spectrum for a Run-2 dataset. The exponent is the usual falling-jet
value~\cite{atlas_dijet_run2}, anywhere from 6 to 8 being defensible; the $r/r_{\mathrm{ref}}$
factor is a sharper
channel's narrower elements holding fewer events each. $W$ is the product over the objects present
of a per-object factor $F$ (Table~\ref{tab:yield}), statistics rather than signal cross section: a
tagged hadronic object costs its mistag rate, a lepton or genuine missing energy the price of an
electroweak process against QCD. Each $F$ is fixed so that Eq.~\eqref{eq:yield} reproduces the
published symmetric channel of its object type to an order of magnitude, the accuracy the
requirement needs; the leptonic $Z$, with no symmetric channel, takes a declared value.

Because $n(m)$ falls monotonically, the populated part of a window is its low-mass end, so the
requirement truncates at the \emph{one-event mass}, where an element drops to a single event, and
discards the spectrum if fewer than 25 elements survive. The one-event mass, the last column of
Table~\ref{tab:yield}, scales as $M_{1} \propto \mathcal{L}^{1/(P-1)}$, which is why a threefold
dataset buys only a fifth in mass reach: on Run~2 and Run~3 combined every entry in that column is
multiplied by $1.20$.

\begin{table}[htbp]
\centering
\caption{The yield model, calibrated object by object. The last two columns follow from
Eq.~\eqref{eq:yield} for the symmetric pair at the Run-2 anchor. Missing energy never enters a mass;
its $F$ prices the category split and is calibrated on $m_{\mathrm{T}}(e\nu)$, the nearest published
channel.}
\label{tab:yield}
\small
\begin{tabular}{llcrrr}
\toprule
object & $F$ & symmetric channel & $r$ & events/element & one-event mass \\
\midrule
light jet & $1$ & $m(jj)$ & 0.050 & $1.0\times10^{6}$ & $10.0\TeV$ \\
$b$-jet & $10^{-1}$ & $m(bb)$ & 0.100 & $2.0\times10^{4}$ & $5.2\TeV$ \\
boosted $W/Z$ & $2\times10^{-2}$ & $m(VV)$ & 0.060 & $4.8\times10^{2}$ & $2.8\TeV$ \\
boosted top & $2\times10^{-2}$ & $m(t\bar t)$ & 0.080 & $6.4\times10^{2}$ & $2.9\TeV$ \\
boosted $H$ & $10^{-2}$ & $m(HH)$ & 0.080 & $1.6\times10^{2}$ & $2.3\TeV$ \\
photon & $4\times10^{-3}$ & $m(\gamma\gamma)$ & 0.010 & $3.2$ & $1.2\TeV$ \\
hadronic $\tau$ & $3.5\times10^{-3}$ & $m(\tau\tau)$ & 0.120 & $29$ & $1.8\TeV$ \\
electron & $3\times10^{-3}$ & $m(ee)$ & 0.015 & $2.7$ & $1.2\TeV$ \\
muon & $3\times10^{-3}$ & $m(\mu\mu)$ & 0.050 & $9.0$ & $1.4\TeV$ \\
$E_{\mathrm{T}}^{\mathrm{miss}}$ & $3\times10^{-3}$ & $m_{\mathrm{T}}(e\nu)$ & 0.100 & $18$ &
  $1.6\TeV$ \\
leptonic $Z$ & $10^{-5}$ & --- & --- & --- & --- \\
\bottomrule
\end{tabular}
\end{table}

A hypothetical spectrum has no published window, so it takes a generic one by multiplicity:
$100$--$5000\GeV$ for a two-body mass, $150$--$4000\GeV$ for a three-body and $200$--$3000\GeV$ for
a
four-body, with the low edge raised to the sum of the object masses (a light object contributing
$40\GeV$) and the high edge truncated as above, so the scanned window is the narrower of the generic
one and what the rate can fill.

\paragraph{The size and direction of the bite.} Of the 21\,644 combinations the ten-object
alphabet allows, 17\,206 fail and 4\,438 survive, the losses concentrated at high multiplicity where
$W$
is a product of many small factors. Four checks bound the consequence. Scaling $N_{\mathrm{ref}}$ by
$10^{2}$ either way, a fair uncertainty for a factorised per-object rate model, moves the
fittable count from 1\,729 to 8\,923 spectra and $\Zloc$ from $6.88$ to $7.14$, inside the band the
mass
resolution already carries. Loosening the two round numbers to 30 events and 15 elements takes
$\Nsig$
to $2.7\times10^{5}$, $+0.04\sigma$, and tightening them to 300 and 50 takes it to
$8.9\times10^{4}$,
$-0.12\sigma$; the element count is the requirement that binds, since at this slope a window holding
25
populated elements holds far more than 100 events. Applied to the published axes, which this study
deliberately avoids, the requirement would keep 49 of the 59 axes this alphabet can form and
$3.3\times10^{3}$ of their $4.2\times10^{3}$ looks, discarding a fifth of the model-space
budget, so exempting them is the conservative choice. And both known biases of the model admit more
spectra rather than fewer: a single power law overestimates yields below a few hundred GeV, where
real
spectra turn over, and it puts the one-event mass of the dijet spectrum at $10\TeV$ against a
published
fit stopping near $8$.

\paragraph{Selection lenses.} A lens keeps the mass axis and its window but costs statistics,
entering as a factor on $W$, the efficiency of its own requirement. Lenses apply one at a time,
never in combination, their objects count against the same four-object ceiling, and every view
faces the statistics requirement, so a thinly populated spectrum affords no lens at all.
Table~\ref{tab:lens} prices the four on the combined dataset, names the physics each is normally
reached for and says which spectra can afford it; statistics rules out 5\,634 of the 9\,407
possible views, the displaced view worst at an assumed efficiency of $10^{-3}$.

The four columns of numbers there are successive filters on the 4\,438 spectra of the unlensed
scan, and the whole estimate is that chain. \emph{Spectra} asks only whether the event has room for
what the lens needs, its objects counting against the same four-object ceiling as the mass: a
high-$H_T$ view needs one object outside the fitted mass, the forward pair two free slots, the ISR
view one free slot and a window reaching below $200\GeV$, and displaced activity nothing extra, so
it reaches every spectrum; summed, 9\,407 conceivable views. \emph{Eff.} is the single number the
lens multiplies into the yield normalisation $W$, the order of magnitude a published search of that
kind retains rather than a measured efficiency; Appendix~\ref{app:uncert} moves all four by a
factor of ten each way and the lensed bar stays inside $7.08$ to $7.15$. \emph{Views} applies the
statistics requirement of Section~\ref{sec:counting} at that reduced $W$, leaving axis, resolution
and window alone except for truncating the ISR view at $200\GeV$: since $n(m) \propto W m^{-6}$, a
tenfold cut in $W$ pulls the one-event mass down by only $10^{1/6} = 1.5$ but divides every element
by ten, so a spectrum near the 100-event floor loses the view outright, hence $3\,453$ of the
$4\,438$ displaced views gone and only $24$ of the $92$ forward ones. \emph{Looks} counts the
survivors with Eq.~\eqref{eq:ns} like any other spectrum.

The $3\,773$ surviving views carry $1.62\times10^{5}$ looks, and
added to the $4\,438$ spectra and $2.01\times10^{5}$ looks of the unlensed scan they give the
$8\,211$ histograms and $\Nsig = 3.6\times10^{5}$ of Table~\ref{tab:granularity}, hence
$\Zloc = 7.11$. No view combines two lenses, so a spectrum gains at most four histograms and on
average $0.85$.

\begin{table}[htbp]
\centering
\caption{The four selection lenses on Run~2 and Run~3 combined, with the model classes of
Appendix~\ref{app:modelmap} that motivate each.}
\label{tab:lens}
\footnotesize
\begin{tabular}{@{}>{\raggedright\arraybackslash}p{0.12\textwidth}%
                  >{\raggedright\arraybackslash}p{0.29\textwidth}%
                  >{\raggedright\arraybackslash}p{0.16\textwidth}rrrr@{}}
\toprule
lens & motivated by & applies when & spectra & eff. & views & looks \\
\midrule
high $H_T$ or $M_\mathrm{eff}$ &
  cascades that leave most of the energy outside the fitted mass: RPV SUSY, pair-produced
  vector-like quarks and leptons, Type-III seesaw, KK gluon &
  an object outside the mass & 4046 & $0.1$ & 2615 & $1.1\times10^{5}$ \\
\addlinespace[2pt]
displaced activity &
  long-lived mediators: the hidden Abelian Higgs at small kinetic mixing, ALPs, prompt-suppressed
  heavy neutral leptons, RPV at small coupling &
  any reconstructed mass & 4438 & $10^{-3}$ & 985 & $3.9\times10^{4}$ \\
\addlinespace[2pt]
forward jet pair &
  electroweak production where gluon fusion is suppressed: heavy 2HDM scalars near alignment,
  Georgi--Machacek and type-II seesaw scalars, KK graviton, HVT &
  two free slots for the tag jets & 92 & $0.02$ & 68 & $3.4\times10^{3}$ \\
\addlinespace[2pt]
ISR jet &
  resonances below the trigger threshold: light $Z'$ and dark photons, dijet-mediator dark matter,
  ALPs, $h\to aa$ in the N2HDM &
  a free slot, and a window below $200\GeV$ & 831 & $0.2$ & 105 & $5.1\times10^{3}$ \\
\bottomrule
\end{tabular}
\end{table}

\input{model_map_appendix}

\section{What a Combinatorial Scan Adds}
\label{app:newspectra}

Section~\ref{sec:scaled} prices a scan over every mass built from at most four objects. Of those
$4\,438$ fittable spectra, this appendix counts how many a public model predicts, and how many of
those no published ATLAS search has scanned.

A spectrum in the scan is one mass observable in one exclusive category, so the test has two parts.
The category has to be a final state some public model produces, and the mass has to sit on the
sub-system of it that resonates. So $m(jj)$ in a two-jet category is the dijet resonance, while the
same $m(jj)$ in a category that also holds two electrons and a photon is a mass no model predicts.
Matching on the mass observable alone would call $70\,\%$ of the scan motivated, which tells us
nothing useful.

The final states are declared per axis from the decay topologies of the classes in
Table~\ref{tab:modelmap}: the resonance's own decay products where it is produced in the $s$
channel, twice those where it is pair-produced and the mass is one leg, and written out for the
cascades ($m(\ell jj)$ inside $\ell\ell jj$, the $bb$ pair inside $4b$, the coloron-pair legs), for
associated production and for vector-boson-fusion tag jets. A predicted final state is the objects
it names and nothing else, so a category that also carries missing energy does not match.

Counted that way, $164$ of the $4\,438$ spectra, or $3.7\,\%$, are predicted. They spread over 59 of
the 63 axes and 76 distinct (final state, mass) pairs, and they carry $\Nsig = 9.2\times10^{3}$, so
scanning only them would cost $\Zloc = 6.58$, a tenth of a sigma above the model-space bar. The
multiplicity above the 76 is the scan's own convention: a four-jet category offers six jet
pairings, each a histogram someone has to fit.

Of those $164$, $51$ sit on an axis no published ATLAS search scans, and Table~\ref{tab:newspectra}
lists them. Together they carry $2\,329$ looks, so opening all of them would move the model-space
bar by well under a tenth of a sigma. What it costs to leave them unscanned is therefore not a
trials factor. It is that nobody has looked.

\begin{table}[htbp]
\centering
\caption{The $51$ spectra of the combinatorial scan that a public model predicts, meaning that the
category is a final state it produces and the mass is the sub-system that resonates, and that no
published ATLAS search scans. They are grouped by the budget axis they sit on. ``spectra'' counts
histograms, so a category offering several object pairings contributes each of them. Model classes
are named as in Table~\ref{tab:modelmap}; the ones cited here are those the literature sweep added,
and the rest come from the FeynRules/UFO database.}
\label{tab:newspectra}
\footnotesize
\input{new_spectra}
\end{table}

\section{The Publication Census}
\label{app:census}

The record behind Section~\ref{sec:published} is the published ATLAS resonance-search literature,
assembled from INSPIRE-HEP and curated by hand: complete to the best of our knowledge, not
machine-verifiable. Its scope matches the budget's, bump hunts for new states, so it excludes
hadron-spectroscopy measurements even where they are bumps in a mass spectrum. Each paper is
attributed to exactly one entry, so the 290 is the sum of the per-entry counts. The two bases are
never summed: the census counts 86 catalogued searches, keeping separate entries for analyses that
share an axis (a high-mass and a trigger-level dijet search, for instance), where the budget merges
them onto one of its 63 axes and counts resolution elements along it.

Two counts separate the 86 entries from the 104 charged pairs of Section~\ref{sec:published}. An
entry may scan more than one axis, and 20 do: the high-mass dilepton search covers $m(ee)$ and
$m(\mu\mu)$, the excited-lepton search $m(e\gamma)$, $m(\mu\gamma)$, $m(ej)$, $m(\mu j)$,
$m(ejj)$ and $m(\tau jj)$, the
single vector-like-quark search three. That makes 114 (search, axis) pairs, over 64 distinct
observables: 44 of the 63 budget axes carry a published search, ten more observables outside the 63
are scanned over a published range, and ten pairs quote no range at all. Seven of the ten have
none published (the three-photon, triboson, high-$p_T$ $Z$-plus-$X$,
triple-Higgs, displaced-diphoton and the two exotic-hadron entries); the other three
declare no single axis to scan (the multilepton and two-body anomaly-detection entries and the
generic multi-body scan). Nothing can be charged for those ten, each a distinct search, so 76 of
the 86 carry looks over 104 pairs. Four of the 104 are charged a single look rather than a window,
their mass fixed and not scanned: $H \to Z\gamma$, the lepton-flavour-violating $Z$ decays,
$\tau \to 3\mu$ and the exclusive quarkonium-plus-photon decays. Of the 104, 29 carry a
model-independent result, judged by the papers' own abstracts: a generic Gaussian-shape limit, a
model-agnostic or anomaly-detection scan, or a cross-section limit the search itself declares model
independent. Those 29 come from 19 of the 76 searches; the other 75 pairs are interpreted only in
specific benchmark models.

Classified by the date of the most recent paper scanning each spectrum, 29 of the 86 are current,
with
a paper from 2024 or later, 38 ageing, between 2019 and 2023, and 19 stale, with nothing since
before
2019. The 6 carrying a published Run-3 result cut across all three.

The repository named in Section~\ref{sec:intro} holds the full list: a per-search table of 86 rows,
reproducing that classification row by row, and all 290 references with title, arXiv number and,
where they exist, journal reference and DOI, grouped under the spectrum each is counted against.

\end{document}

%% file: two_body_matrix.tex
\begin{tabular}{lrrrrrrrrr}
\toprule
 & $e$ & $\mu$ & $\tau$ & $q/g$ & $b$ & $t$ & $\gamma$ & $Z/W$ & $H$ \\
\midrule
$e$ & 630 & 221 & 28 & 107 & \textit{(58)} & \textit{(29)} & 215 & 80 & \textit{(25)} \\
$\mu$ &  & 486 & 28 & 80 & \textit{(46)} & \textit{(29)} & 107 & 60 & \textit{(25)} \\
$\tau$ &  &  & 43 & 32 & 19 & \textit{(31)} & \textit{(40)} & \textit{(24)} & \textit{(30)} \\
$q/g$ &  &  &  & 74 & 57 & 27 & 66 & \textit{(44)} & \textit{(25)} \\
$b$ &  &  &  &  & 40 & 44 & \textit{(64)} & 63 & \textit{(39)} \\
$t$ &  &  &  &  &  & 36 & \textit{(38)} & 40 & 14 \\
$\gamma$ &  &  &  &  &  &  & 402 & 110 & \textit{(52)} \\
$Z/W$ &  &  &  &  &  &  &  & 57 & 54 \\
$H$ &  &  &  &  &  &  &  &  & 40 \\
\bottomrule
\end{tabular}

%% file: resolution_appendix.tex
\section{The Origin of the Fractional Resolutions}
\label{app:resolution}

The fractional mass resolution $r$ is the one genuine physics input to the budget, entering
Eq.~\eqref{eq:ns} as $n_s \propto 1/r$, so it sets the trials count directly. This appendix says
where each value comes from and how far to trust it.

\paragraph{The status of these numbers.} A published search quotes a mass resolution for its own
selection, range and calibration; no uniform set of 63 such numbers exists, and inventing one per
spectrum would feign precision. So \budgetvalues{} are \emph{estimates propagated from ATLAS
object-performance measurements}: for each spectrum, the fractional momentum or energy resolution
ATLAS has published for the object that limits it. Table~\ref{tab:resolution} lists those objects
and measurements; every $r$ in the budget is one of them, or a combination for a multi-object mass.

\begin{table}[htbp]
\centering
\small
\caption{The object performance behind the fractional resolutions: the $r$ assigned to a two-body mass
limited by that object, and the ATLAS measurement it is propagated from.}
\label{tab:resolution}
\begin{tabular}{@{}l>{\raggedright\arraybackslash}p{0.42\textwidth}cl@{}}
\toprule
limiting object & what sets the resolution & $r$ & from \\
\midrule
$e$, $\gamma$ & electromagnetic calorimeter energy resolution, nearly
  mass-independent & $0.010$--$0.015$ & \cite{atlas_egamma_calib} \\
$\mu$ (high mass) & sagitta of a curved track; the fractional resolution grows with $p_{\mathrm{T}}$
  & $0.05$ & \cite{atlas_muon_perf} \\
$\mu$ (low mass) & tracker-dominated, no sagitta penalty at low $p_{\mathrm{T}}$ & $0.02$ &
  \cite{atlas_muon_perf} \\
light jet & jet energy scale and resolution & $0.05$ & \cite{atlas_jer} \\
$b$-jet & as a light jet, degraded by the neutrinos of semileptonic $b$ decays & $0.06$--$0.10$ &
  \cite{atlas_jer, atlas_btag_perf} \\
$\tau_{\mathrm{had}}$ & visible decay products only; the neutrino is never reconstructed & $0.10$--$0.12$ &
  \cite{atlas_tau_reco} \\
boosted $V$, $h$, $t$, $H$ & large-radius jet mass, calibrated as one object & $0.06$--$0.08$ &
  \cite{atlas_largeR_mass} \\
$E_{\mathrm{T}}^{\mathrm{miss}}$ (transverse masses) & soft-term and pile-up dependent, the coarsest
  axis in the budget & $0.10$--$0.15$ & \cite{atlas_met_perf} \\
\bottomrule
\end{tabular}
\end{table}

\paragraph{From objects to spectra.} A two-body mass takes its worse leg, which dominates the
quadrature sum: $m(e\mu)$ sits at $0.03$, not the electron's $0.015$, and $m(\tau b)$ at $0.12$,
not the $b$-jet's $0.06$. Composite objects are assigned their large-radius jet mass resolution
rather than that of their decay products, which is why $m(VV)$, $m(HH)$, $m(t\bar t)$ and the other
boosted axes cluster at $0.06$--$0.08$. Transverse masses, inheriting the missing-momentum
resolution and peakless on the high side, are the coarsest axes in the program.

\paragraph{The one place a constant $r$ is a real approximation.} Equation~\eqref{eq:ns} assumes a
constant $r$, true for electrons and photons and false for muons, whose sagitta resolution runs
from roughly $2\,\%$ at $200\GeV$ to $\sim15\,\%$ at $3\TeV$. The muon channels therefore take an
\emph{effective flat} $r$, the constant reproducing the trials integral of the true rising $r(M)$
over that channel's own scan window: for $m(\mu\mu)$ over $150\GeV$ to $8\TeV$ the integral gives
$n_s = 76$, hence $r_{\mathrm{eff}} = 0.05$. This is a window-averaged stand-in, not a measurement,
a mass-dependent $r(M)$ being the natural next refinement, and the largest single modelling
approximation in the budget.

At low mass the ordering of the two lepton flavours reverses, stated here because it looks like an
error otherwise: a few-GeV muon pair is a sharp tracker measurement while an $e^+e^-$ pair of the
same mass is a marginally resolved calorimeter cluster, so the dark-photon muon axis carries
$r = 0.02$ against the electron axis at $0.025$, the opposite of the high-mass pattern.

\paragraph{The narrow-resonance assumption behind $r$.} Equation~\eqref{eq:ns} counts
\emph{detector} resolution elements, the correct step size only while the signal's natural width
stays below $r$: a broader resonance correlates neighbouring mass points, leaving fewer independent
looks. The direction is what matters: counting elements for a wide signal \emph{over}-counts
$\Nsig$ and makes $\Zloc$ too strict, never too loose, so no width correction can lower the
discovery bar quoted in Section~\ref{sec:budget}.

Of the 58 public model classes behind the 63 axes, 36 are narrow on every axis they populate:
$\Gamma/M \sim \varepsilon^{2}$ for a kinetically mixed dark photon, $\lambda^{2}/16\pi \approx
2\,\%$ for a leptoquark at $\lambda = 1$, $2$--$3\,\%$ for a $W_R$, $2$--$4\,\%$ for an excited
quark. Sixteen are narrow only at the benchmark ATLAS publishes and broad elsewhere in their parameter
space: a $Z'_{\mathrm{SSM}}$ carries $\Gamma/M \approx 3\,\%$ against $r = 0.015$ on
$m(ee)$, the simplified dark-matter mediator spans $1\,\%$ to over $30\,\%$ across its coupling
grid, single vector-like quark production reaches $10$ to $50\,\%$, and the vector leptoquark $U_1$
sits near $20\,\%$ at the coupling that fits the flavour anomalies. Three are already wider than
$r$ at the standard benchmark, the Kaluza--Klein gluon worst at $15$ to $30\,\%$ against
$r = 0.08$ on $m(t\bar t)$, followed by the coloron/axigluon and composite classes. The remaining
three have no Breit--Wigner peak at all: a quantum black hole is a threshold turn-on, ADD and HEIDI
give a non-resonant high-mass tail, and toponium is a threshold effect pinned at $2m_t$ rather than
a scannable mass.

Peaking belongs to the model \emph{and} the axis, so further cases are recorded against
specific axes: prompt heavy neutrinos give a genuine $m(\ell\ell jj)$ resonance but only a counting
signature in multilepton, Type-III seesaw triplets and vector-like leptons peak on $m(\ell Z)$ but
are pair-produced counting signatures in multilepton, and heavy two-Higgs-doublet scalars interfere
with the Standard Model
$t\bar t$ continuum, a peak-dip lineshape on $m(t\bar t)$ rather than a bump. Only one of the 63
axes is motivated \emph{exclusively} by non-peaking models, $m(\mathrm{multi})$,
and dropping it takes $\Nsig$ from 4319 to 4298 and $\Zloc$ from $6.46$ to $6.46$: a thousandth
of a sigma. The combinatorial scan is insulated by construction, the four axes
with no scannable object composition, $m(\mathrm{multi})$ and the three transverse masses, being
exactly the observables it cannot form.

\paragraph{The size of the effect on the answer.} Every headline carries the band from scaling $r$
by two each way, so the sensitivity is already in the result: a factor two in $r$ is a factor two
in $\Nsig$ and about $0.1\sigma$ on $\Zloc$. That band is far wider than the gap between any two
defensible choices of these values, which is why a coarse per-object estimate suffices here; it
would not for a limit, and nothing here should be read as a resolution measurement.

%% file: uncertainty_table.tex
\begin{tabular}{@{}l>{\raggedright\arraybackslash}p{0.30\textwidth}ccc@{}}
\toprule
source & varied over & 63 spectra & scan & difference \\
\midrule
mass resolution, scale & every $r \times 2$ either way & $^{+0.11}_{-0.11}$ & $^{+0.13}_{-0.19}$ & $^{+0.03}_{-0.08}$ \\
mass resolution, per channel$^{\dagger}$ & each $r$ independently, $\times 2$ per $\sigma$ (16--84\%) & $^{+0.02}_{-0.02}$ & --- & $^{+0.02}_{-0.02}$ \\
mass resolution, shape & muon axes, $r(M)$ rising to 0.10--0.20 at 3\,TeV & $<0.01$ & --- & $<0.01$ \\
mass resolution, prescription & worst leg or quadrature sum, not the calibrated mean & --- & $-0.10$ & $-0.10$ \\
scan windows & every edge $\times 1.4$ (published), $\times 1.25$ (generic) & $^{+0.03}_{-0.04}$ & $^{+0.02}_{-0.02}$ & $^{+0.02}_{-0.01}$ \\
yield anchor & $N_{\mathrm{ref}} \times 10^{\pm2}$ & --- & $^{+0.11}_{-0.15}$ & $^{+0.11}_{-0.15}$ \\
background slope & $P = 6$ to $8$ & --- & $^{+0.01}_{-0.01}$ & $^{+0.01}_{-0.01}$ \\
fittability requirement & 30--300 events, 15--50 elements & --- & $^{+0.04}_{-0.12}$ & $^{+0.04}_{-0.12}$ \\
the axis set & non-peaking axes out, the 4 unscanned pairs in & $+0.01$ & --- & $-0.01$ \\
the definition of one look & $N \times 0.5$ to $N Z/\sqrt{2\pi}$ & $^{+0.14}_{-0.11}$ & $^{+0.15}_{-0.10}$ & $+0.01$ \\
the closed-form LEE relation & exact Gaussian-tail solution & $-0.04$ & $-0.05$ & $-0.01$ \\
\midrule
total & in quadrature & $^{+0.18}_{-0.16}$ & $^{+0.23}_{-0.31}$ & $^{+0.12}_{-0.23}$ \\
\bottomrule
\end{tabular}

%% file: model_map_appendix.tex
\section{The Model-to-Spectrum Map}
\label{app:modelmap}

Table~\ref{tab:modelmap} is the input the count of Section~\ref{sec:budget} is built
from: each canonical spectrum the budget scans, the public model classes predicting a resonance in
it, and the published event selections that scan it. Rows follow Figure~\ref{fig:budget}, most
looks first. The model classes come from the public FeynRules
database~\cite{feynrules,feynrules_db}, which distributes BSM Lagrangians in the UFO
format~\cite{ufo} that the LHC generators read, together with the published search record and a
sweep of the resonance-model literature beyond the database: the 15 classes the
database does not implement carry their defining references in the table, and a class with no
reference behind it comes from the database, covered by that single
reference~\cite{feynrules_db}. One
entry is a class of models sharing a decay topology rather than a single Lagrangian, so
\emph{2HDM} covers the general, type-II and CP-violating implementations together: two
implementations peaking in the same spectrum are tested by the same search. Classes whose only
signature is non-resonant (mono-$X$, anomalous $\mathrm{d}E/\mathrm{d}x$, displaced-only decays)
are absent, populating no mass spectrum and buying no trials. The map is many-to-many, and its
off-diagonal weight is what makes breadth cheap: 58 classes make 211
class--spectrum pairs over 63 spectra, 3.3 per spectrum on average and
13 on $m(VV)$.

A channel $c_s$ is a distinct event selection producing its own bump spectrum: a $b$-tag category, a
boost regime, a sub-decay mode, or an ambiguous object pairing the search histograms both ways. A fit
category combined into a single limit does not count, and $c_s$ is a multiplicity \emph{within} one
spectrum, never reaching across to a different final state. That column sums to 111 and
$\Nsig = \sum_s c_s n_s = 7\,231$, the selections-level entry of
Table~\ref{tab:granularity}. If a search family does combine its channels, counting each as an
independent scan over-counts the trials, which is why that level is an upper bracket on the inclusive
count rather than a replacement for it.

\begingroup\small\setlength{\tabcolsep}{4pt}
\begin{longtable}{@{}>{\raggedright\arraybackslash}p{0.145\textwidth}rrcc%
>{\raggedright\arraybackslash}p{0.315\textwidth}%
>{\raggedright\arraybackslash}p{0.28\textwidth}@{}}
\caption{The 63 spectra of the budget, in the order of Figure~\ref{fig:budget}: the public
model classes predicting a resonance in each, and the published event selections counted against
it. The pub.\ column marks the 44 axes at least one published ATLAS search already scans
(Appendix~\ref{app:census}); the scan column marks the 59 the combinatorial catalogue of
Section~\ref{sec:scaled} can form, the other 4 needing missing energy in
the mass or, for $m(\mathrm{multi})$, having no fixed object composition.}\label{tab:modelmap}\\
\toprule
spectrum & $n_s$ & $c_s$ & pub. & scan & public model classes & event selections \\
\midrule
\endfirsthead
\caption[]{\emph{continued.}}\\
\toprule
spectrum & $n_s$ & $c_s$ & pub. & scan & public model classes & event selections \\
\midrule
\endhead
\midrule
\multicolumn{2}{@{}l}{total} & 111 & & & & \\
\bottomrule
\endlastfoot
$m(\gamma\gamma)$ & 402 & 2 & yes & yes & ALP, KK graviton (Gstar), Large ED / UED / HEIDI, N2HDM / 2HDM+S (h$\to$aa), Randall-Sundrum / Radion, Singlet scalar / SM+Scalars, Vector-like confinement & spin-0 (ggF) + spin-2/VBF (converted/unconverted categories) \\
$m(\mu\mu)$ (Zd) & 360 & 1 & yes & yes & Dark Z (mass mixing)~\cite{lit:1203.2947,lit:1412.0018}, Hidden Abelian Higgs (HAHM), N2HDM / 2HDM+S (h$\to$aa) & prompt muon lepton-jet / $4\mu$ scan \\
$m(ee)$ & 265 & 1 & yes & yes & KK graviton (Gstar), Large ED / UED / HEIDI, Leptophilic gauge boson, LRSM / Alt-LRSM, Minimal Z' / U(1), SILH / Little Higgs, Technicolor / TC2 & single dielectron scan (barrel/endcap are fit categories, combined) \\
$m(ee)$ (Zd) & 240 & 1 & yes & yes & Dark Z (mass mixing)~\cite{lit:1203.2947,lit:1412.0018}, Hidden Abelian Higgs (HAHM) & prompt $e$ lepton-jet / $4e$ scan \\
$m(e\gamma)$ & 215 & 1 & yes & yes & Excited lepton (l*) & excited electron $e^*$ \\
$m(e\mu)$ LFV & 146 & 1 & yes & yes & Minimal Z' / U(1), Quantum black hole, RPV resonant slepton~\cite{lit:hep-ph/0001224,lit:1201.5014}, Scalar leptoquark (S1), Vector leptoquark (U1) & LFV $e\mu$ \\
$m(ee)$ SS & 125 & 1 & yes & yes & Bilepton (331)~\cite{lit:1806.04536,lit:1812.02723}, Georgi-Machacek, Type-II seesaw, Zee-Babu~\cite{lit:1402.4491,lit:2206.14833} & $H^{++}$ same-sign $ee$ \\
$m(V\gamma)$ & 110 & 2 & yes & yes & ALP, Excited boson (W*/Z*), Singlet scalar / SM+Scalars & $X \to Z\gamma$: $Z \to \ell\ell$ and boosted $Z \to qq$ (+ $W\gamma$) \\
$m(ej)$ & 107 & 4 & yes & yes & Leptogluon (color-octet l)~\cite{lit:1211.6394,lit:2212.06178}, Leptoquark NLO (mix/nomix), Quantum black hole, Scalar leptoquark (S1) & LQ pair $\to eq$: all four lepton--jet pairings scanned (arXiv:2006.05872) \\
$m(\mu \gamma)$ & 107 & 1 & yes & yes & Excited lepton (l*) & excited muon $\mu^*$ \\
$m(\mu j)$ & 80 & 4 & yes & yes & Leptogluon (color-octet l)~\cite{lit:1211.6394,lit:2212.06178}, Leptoquark NLO (mix/nomix), Quantum black hole, Scalar leptoquark (S1) & LQ pair $\to \mu q$: all four lepton--jet pairings scanned (arXiv:2006.05872) \\
$m(eZ)$ & 80 & 1 & yes & yes & Excited lepton (l*), RPV electroweakino (trilepton), Type-III seesaw, Vector-like lepton (VLL) & trilepton $e+Z$ resonance \\
$m(\mu\mu)$ & 80 & 1 & yes & yes & B-anomaly Z' (b-philic)~\cite{lit:1809.01158,lit:1403.1269,lit:1707.07016}, KK graviton (Gstar), Large ED / UED / HEIDI, Leptophilic gauge boson, LRSM / Alt-LRSM, Minimal Z' / U(1), SILH / Little Higgs, Technicolor / TC2 & single dimuon scan (combined charge/$\eta$ categories) \\
$m(cb)$ dijet & 78 & 2 & -- & yes & 2HDM (general/typeII/CPV) & light $H^+ \to cb$ in $t\bar t$: hadronic + semileptonic tag \\
$m(e\mu)$ SS & 75 & 1 & yes & yes & Bilepton (331)~\cite{lit:1806.04536,lit:1812.02723}, Georgi-Machacek, Type-II seesaw, Zee-Babu~\cite{lit:1402.4491,lit:2206.14833} & $H^{++}$ same-sign $e\mu$ \\
$m(jj)$ & 74 & 2 & yes & yes & Color-octet scalar (MW)~\cite{lit:hep-ph/0606172,lit:0710.3133}, Coloron / Axigluon, Composite / NJL, Diquark / color-sextet, Excited quark (q*/b*), KK graviton (Gstar), Minimal Z' / U(1), Simplified DM (dijet mediator), String / Regge resonance~\cite{lit:0808.0497,lit:0804.2013}, Vector-like confinement, W' & low-mass (TLA/ISR) + high-mass inclusive dijet \\
$m(j\gamma)$ & 66 & 1 & yes & yes & Excited quark (q*/b*), Quantum black hole, String / Regge resonance~\cite{lit:0808.0497,lit:0804.2013} & $q^* \to q\gamma$ (single photon+jet SR) \\
$m(b\gamma)$ & 64 & 1 & -- & yes & Excited quark (q*/b*) & $b^* \to b\gamma$: no ATLAS search at any energy \\
$m(ejj)$ & 64 & 1 & yes & yes & Composite Majorana N~\cite{lit:1510.07988}, Excited lepton (l*), Excited neutrino (nu*)~\cite{lit:hep-ph/0401066,lit:2606.24486}, Heavy neutrino / HNL (prompt), LRSM / Alt-LRSM & $e^*/\nu^* \to e\,qq$ (the $e^*$ system of the $eejj$ search) \\
$m(\mu Z)$ & 60 & 1 & yes & yes & Excited lepton (l*), RPV electroweakino (trilepton), Type-III seesaw, Vector-like lepton (VLL) & trilepton $\mu+Z$ resonance \\
$m(\gamma jj)$ & 60 & 1 & -- & yes & Technicolor / TC2, Warped KK cascade~\cite{lit:1612.00047,lit:1711.09920} & $\gamma$ + dijet cascade: no search at any collider \\
$m(eb)$ & 58 & 2 & -- & yes & Leptoquark NLO (mix/nomix), Scalar leptoquark (S1), Stop/scharm (RPV) & RPV stop $\to be$ + $b$-tagged LQ leg (both pairings) \\
$m(eejj)$ & 57 & 1 & yes & yes & Heavy neutrino / HNL (prompt), LRSM / Alt-LRSM & $W_R \to eejj$ \\
$m(VV)$ & 57 & 9 & yes & yes & Composite / NJL, Excited boson (W*/Z*), Georgi-Machacek, HeavyHiggs THDM, KK graviton (Gstar), Randall-Sundrum / Radion, SILH / Little Higgs, Singlet scalar / SM+Scalars, Technicolor / TC2, TRSM (singlet/triplet), Type-II seesaw, Vector triplet (HVT), Warped KK cascade~\cite{lit:1612.00047,lit:1711.09920} & $WW/WZ/ZZ$ $\times$ $qqqq$ / $\ell\nu qq$ / $\ell\ell qq$ / $\nu\nu qq$ / $\ell\nu\ell\nu$, ggF+VBF \\
$m(bj)$ & 57 & 2 & yes & yes & Excited quark (q*/b*) & $b^* \to bg$: $b$-tagged leading + subleading jet pairing \\
$m(Vh)$ & 54 & 6 & yes & yes & 2HDM (general/typeII/CPV), Vector triplet (HVT) & $Wh/Zh$ $\times$ 0/1/2-lepton $\times$ $h \to bb$ resolved/boosted \\
$m(\mu\mu)$ SS & 47 & 1 & yes & yes & Bilepton (331)~\cite{lit:1806.04536,lit:1812.02723}, Georgi-Machacek, Type-II seesaw, Zee-Babu~\cite{lit:1402.4491,lit:2206.14833} & $H^{++}$ same-sign $\mu\mu$ \\
$m(\mu b)$ & 46 & 2 & -- & yes & Leptoquark NLO (mix/nomix), Scalar leptoquark (S1), Stop/scharm (RPV) & RPV stop $\to b\mu$ + $b$-tagged LQ leg (both pairings) \\
$m(bZ)$ & 46 & 1 & yes & yes & Vector-like quark (VLQ) & VLQ $b'/B \to bZ(\ell\ell)$ (7/8 TeV scans only) \\
$m(\mu j j)$ & 45 & 1 & -- & yes & Composite Majorana N~\cite{lit:1510.07988}, Excited lepton (l*), Excited neutrino (nu*)~\cite{lit:hep-ph/0401066,lit:2606.24486}, Heavy neutrino / HNL (prompt), LRSM / Alt-LRSM, RPV resonant slepton~\cite{lit:hep-ph/0001224,lit:1201.5014} & $\mu^*/\nu^* \to \mu\,qq$: no ATLAS search \\
multilepton & 44 & 4 & -- & yes & Heavy neutrino / HNL (prompt), Leptophobic Z' (dark cascade)~\cite{lit:1111.0633}, Type-III seesaw, Vector-like lepton (VLL) & $3\ell$ / $4\ell$ signal regions by flavour and charge (axis deliberately unsplit) \\
$m(jV)$ & 44 & 1 & -- & yes & Excited quark (q*/b*), Vector-like quark (VLQ) & $q^* \to qW/qZ$: no ATLAS search of the gauge decays \\
$m(3j)$ & 44 & 1 & yes & yes & MSSM/NMSSM/RPV SUSY, Sgluon, Warped KK cascade~\cite{lit:1612.00047,lit:1711.09920} & RPV/sgluon three-jet resonance (single SR) \\
$m(tb)$ & 44 & 3 & yes & yes & 2HDM (general/typeII/CPV), Vector-like quark (VLQ), W' & $W'/H^+ \to tb$: 0-lepton, 1-lepton $\times$ $b$-tag \\
$m(\tau\tau)$ & 43 & 3 & yes & yes & 2HDM (general/typeII/CPV), Minimal Z' / U(1), MSSM/NMSSM/RPV SUSY, N2HDM / 2HDM+S (h$\to$aa), Vector leptoquark (U1) & $H/Z' \to \tau\tau$: $\tau_e\tau_h$, $\tau_\mu\tau_h$, $\tau_h\tau_h$ \\
$m(\mu\mu jj)$ & 41 & 1 & yes & yes & Heavy neutrino / HNL (prompt), LRSM / Alt-LRSM, RPV resonant slepton~\cite{lit:hep-ph/0001224,lit:1201.5014} & $W_R \to \mu\mu jj$ \\
$m(\tau \gamma)$ & 40 & 1 & -- & yes & Excited lepton (l*) & $\tau^* \to \tau\gamma$: no ATLAS search at any energy \\
$m(bb)$ & 40 & 3 & yes & yes & 2HDM (general/typeII/CPV), B-anomaly Z' (b-philic)~\cite{lit:1809.01158,lit:1403.1269,lit:1707.07016}, Color-octet scalar (MW)~\cite{lit:hep-ph/0606172,lit:0710.3133}, Minimal Z' / U(1), MSSM/NMSSM/RPV SUSY, N2HDM / 2HDM+S (h$\to$aa), TRSM (singlet/triplet) & $h \to aa \to 4b$ low-mass + $A/Y \to bb$ resolved + boosted \\
$m(HH)$ & 40 & 6 & yes & yes & 2HDM (general/typeII/CPV), Higgs portal, KK graviton (Gstar), MSSM/NMSSM/RPV SUSY, Randall-Sundrum / Radion, Singlet scalar / SM+Scalars, TRSM (singlet/triplet) & $bbbb$ resolved + boosted, $bb\tau\tau$, $bb\gamma\gamma$, $bbVV$, multilepton \\
$m_{\mathrm{T}}(e\nu)$ & 38 & 1 & yes & -- & LRSM / Alt-LRSM, W' & $W' \to e\nu$ \\
$m(t\gamma)$ & 38 & 1 & -- & yes & Excited top (t*)~\cite{lit:1208.5811,lit:1110.1565} & $t^* \to t\gamma$: no ATLAS search \\
$m(tt)$ & 36 & 4 & yes & yes & 2HDM (general/typeII/CPV), Coloron / Axigluon, Diquark / color-sextet, HeavyHiggs THDM, KK gluon (RS), Minimal Z' / U(1), Sgluon, Top-philic, Toponium & $t\bar t$ resonance: $\ell$+jets resolved, $\ell$+jets boosted, all-hadronic, dilepton \\
$m(\tau j)$ & 32 & 2 & yes & yes & Leptoquark NLO (mix/nomix), Scalar leptoquark (S1), Vector leptoquark (U1) & LQ $\to \tau q$: $\tau_{\mathrm{had}}$ + $\tau_{\mathrm{lep}}$ selections \\
$m(et)$ & 29 & 1 & -- & yes & Scalar leptoquark (S1), Vector leptoquark (U1) & LQ $\to te$: no ATLAS search \\
$m(\mu t)$ & 29 & 1 & -- & yes & Scalar leptoquark (S1), Vector leptoquark (U1) & LQ $\to t\mu$: no ATLAS search \\
$m(e\tau)$ LFV & 28 & 1 & yes & yes & Minimal Z' / U(1), Quantum black hole, RPV resonant slepton~\cite{lit:hep-ph/0001224,lit:1201.5014}, Scalar leptoquark (S1), Vector leptoquark (U1) & LFV $e\tau$ (hadronic $\tau$) \\
$m(\mu\tau)$ LFV & 28 & 1 & yes & yes & Minimal Z' / U(1), Quantum black hole, RPV resonant slepton~\cite{lit:hep-ph/0001224,lit:1201.5014}, Scalar leptoquark (S1), Vector leptoquark (U1) & LFV $\mu\tau$ (hadronic $\tau$) \\
$m(\tau jj)$ & 28 & 1 & yes & yes & Excited lepton (l*) & $\tau^* \to \tau\,qq$ (contact interaction) \\
$m(tj)$ & 27 & 1 & yes & yes & Excited top (t*)~\cite{lit:1208.5811,lit:1110.1565}, Flavoured W'/Z' (t-q)~\cite{lit:1102.0018,lit:0907.4112} & $W'/Z' \to t\,+$ jet (7 TeV scan only) \\
$m_{\mathrm{T}}(\tau\nu)$ & 27 & 1 & yes & -- & 2HDM (general/typeII/CPV), W' & $W' \to \tau\nu$ (single hadronic-$\tau$ channel) \\
$m_{\mathrm{T}}(\mu\nu)$ & 26 & 1 & yes & -- & LRSM / Alt-LRSM, W' & $W' \to \mu\nu$ \\
$m(eH)$ & 25 & 1 & -- & yes & Type-III seesaw, Vector-like lepton (VLL) & $L/\Sigma \to e\,h$: no ATLAS search \\
$m(\mu H)$ & 25 & 1 & -- & yes & Type-III seesaw, Vector-like lepton (VLL) & $L/\Sigma \to \mu\,h$: no ATLAS search \\
$m(jH)$ & 25 & 1 & -- & yes & Vector-like quark (VLQ) & light-flavour VLQ $\to qh$: no search at any collider \\
$m(\tau V)$ & 24 & 1 & -- & yes & Excited lepton (l*), Type-III seesaw, Vector-like lepton (VLL) & $\tau^*/L/\Sigma \to \tau Z/W$: no ATLAS search \\
$m(tW)$ & 22 & 2 & yes & yes & Excited quark (q*/b*), Vector-like quark (VLQ) & $b^* \to tW$: leptonic + hadronic top \\
$m(\mathrm{multi})$ & 21 & 1 & -- & -- & Large ED / UED / HEIDI, Quantum black hole & QBH / multijet (single high-multiplicity SR) \\
$m(tbj)$ & 20 & 1 & -- & yes & MFV RPV gluino (tbs)~\cite{lit:1111.1239} & MFV RPV gluino $\to tbs$ (no mass scan anywhere) \\
$m(tt)/m(jj)$ & 20 & 2 & -- & yes & Coloron / Axigluon, Stop/scharm (RPV) & coloron pair $\to tt$ or $jj$ decay legs; the scan targets the leg mass as $m(jj)$ in the $4j$ and $2t2j$ categories and as $m(tt)$ in $2t2j$ (the all-top pair category is too thin to fit) \\
$m(\tau b)$ & 19 & 2 & yes & yes & Leptoquark NLO (mix/nomix), Scalar leptoquark (S1), Vector leptoquark (U1) & LQ$_3 \to \tau b$: $\tau_{\mathrm{had}}$ + $\tau_{\mathrm{lep}}$ \\
$m(ttZ)/m(Zt)$ & 17 & 2 & yes & yes & Vector-like quark (VLQ) & VLQ single/pair $T \to tZ$ selections \\
$m(Wb)$ & 17 & 1 & yes & yes & Vector-like quark (VLQ) & single VLQ $T/Y \to Wb$ (1-lepton SR) \\
$m(Ht)$ & 14 & 1 & yes & yes & Vector-like quark (VLQ) & single VLQ $T \to Ht$ ($h \to bb$ tagged) \\
\end{longtable}
\endgroup

%% file: new_spectra.tex
\begin{tabular}{@{}l>{\raggedright\arraybackslash}p{0.14\textwidth}>{\raggedright\arraybackslash}p{0.10\textwidth}rr>{\raggedright\arraybackslash}p{0.30\textwidth}@{}}
\toprule
axis & mass & final state(s) & spectra & looks & model classes that predict it \\
\midrule
$m(tt)/m(jj)$ & $m(2j)$, $m(2t)$ & $2j2t$, $4j$ & 8 & 568 & Coloron / Axigluon, Stop/scharm (RPV) \\
$m(eb)$ & $m(eb)$ & $2e2b$, $eb$ & 9 & 289 & Leptoquark NLO (mix/nomix), Scalar leptoquark (S1), Stop/scharm (RPV) \\
$m(\mu b)$ & $m({\mu}b)$ & $2{\mu}2b$, ${\mu}b$ & 9 & 263 & Leptoquark NLO (mix/nomix), Scalar leptoquark (S1), Stop/scharm (RPV) \\
$m(\mu j j)$ & $m({\mu}2j)$ & $2{\mu}2j$, ${\mu}2j$ & 5 & 262 & Composite Majorana N~\cite{lit:1510.07988}, Excited lepton (l*), Excited neutrino (nu*)~\cite{lit:hep-ph/0401066,lit:2606.24486}, Heavy neutrino / HNL (prompt), LRSM / Alt-LRSM, RPV resonant slepton~\cite{lit:hep-ph/0001224,lit:1201.5014} \\
$m(cb)$ dijet & $m(jb)$ & $2j2b$ & 4 & 198 & 2HDM (general/typeII/CPV) \\
multilepton & $m(2e{\mu})$, $m(3e)$, $m(3{\mu})$, $m(e2{\mu})$ & $2e{\mu}$, $3e$, $3{\mu}$, $e2{\mu}$ & 4 & 194 & Heavy neutrino / HNL (prompt), Leptophobic Z' (dark cascade)~\cite{lit:1111.0633}, Type-III seesaw, Vector-like lepton (VLL) \\
$m(\gamma jj)$ & $m({\gamma}2j)$ & ${\gamma}2j$ & 1 & 80 & Technicolor / TC2, Warped KK cascade~\cite{lit:1612.00047,lit:1711.09920} \\
$m(jV)$ & $m(jV)$ & $jV$ & 1 & 66 & Excited quark (q*/b*), Vector-like quark (VLQ) \\
$m(jH)$ & $m(jH)$ & $jH$ & 1 & 51 & Vector-like quark (VLQ) \\
$m(b\gamma)$ & $m({\gamma}b)$ & ${\gamma}b$ & 1 & 50 & Excited quark (q*/b*) \\
$m(eH)$ & $m(eH)$ & $eH$ & 1 & 45 & Type-III seesaw, Vector-like lepton (VLL) \\
$m(t\gamma)$ & $m({\gamma}t)$ & ${\gamma}t$ & 1 & 44 & Excited top (t*)~\cite{lit:1208.5811,lit:1110.1565} \\
$m(et)$ & $m(et)$ & $et$ & 1 & 42 & Scalar leptoquark (S1), Vector leptoquark (U1) \\
$m(\mu H)$ & $m({\mu}H)$ & ${\mu}H$ & 1 & 39 & Type-III seesaw, Vector-like lepton (VLL) \\
$m(\mu t)$ & $m({\mu}t)$ & ${\mu}t$ & 1 & 37 & Scalar leptoquark (S1), Vector leptoquark (U1) \\
$m(\tau \gamma)$ & $m({\tau}{\gamma})$ & ${\tau}{\gamma}$ & 1 & 35 & Excited lepton (l*) \\
$m(tbj)$ & $m(jbt)$ & $jbt$ & 1 & 35 & MFV RPV gluino (tbs)~\cite{lit:1111.1239} \\
$m(\tau V)$ & $m({\tau}V)$ & ${\tau}V$ & 1 & 32 & Excited lepton (l*), Type-III seesaw, Vector-like lepton (VLL) \\
\midrule
total & & & 51 & 2329 & \\
\bottomrule
\end{tabular}